\documentstyle[epsfig]{aipproc2}
\begin{document}
\title{Primordial molecular clouds}

\author{Denis Puy$^*$ and Monique Signore$^{\dagger}$}
\address{$^*$Paul Scherrer Institute, Villigen (Switzerland)\\
Institute of Theoretical Physics, University of Zurich (Switzerland)
\vskip2mm
$^{\dagger}$ Observatoire de Paris, Laboratoire de Radioastronomie (France)}

\maketitle

\begin{abstract}
It is now well known that a primordial chemistry, involving light elements
produced during the nucleosynthesis period, might develop during the hydrogen
post-recombination era. In particular, molecular ions and primordial molecules
such as $H_2$, $HD$ and $LiH$ will be produced. We summarize this primordial
chemistry after the recombination epoch, and then present a simple gravitational
collapse model of a cloud. The potentiality of fragmentation of this collapsing
protoclouds through the thermal instability is also discussed. We suggest that
this study could also be extended to the $CO$ molecule, because the carbon
reservoir molecule $CO$ has already been observed in high redshifts objects.
\end{abstract}

\section{Introduction}
The study of primordial 
chemistry of molecules addresses a number of interesting questions
pertaining to the thermal balance of collapsing molecular
protoclouds. 
In numerous astrophysical cases molecular cooling and heating influence 
dynamical evolution of the medium, see for example the review done 
by Dalgarno and Mc Cray (1972) who 
present molecular cooling and heating in HI regions, or Shu (1997) 
who discusses the influence of molecules in star formation.
\\
The existence of a
significant abundance of molecules can be crucial on the dynamical
evolution of collapsing objects. Because the cloud temperature 
increases with contraction, a cooling mechanism can be important for
the structure formation, 
by lowering pressure opposing gravity, i.e. by allowing
continued collapse of Jeans unstable protoclouds. This is particularly
true for the first generation of objects. Many authors have
examined this problem (Bodenheimer 1968, Hutchins 1976, Palla et al
1983, Villere \& Bodenheimer 1987, 
Tegmark et al 1997) introducing molecular coolants, mainly $H_2$ molecules.
From Fall and Rees (1985) we know
that $H_2$ molecules are important cooling agents and can lead,
sometimes, to the fragmentation process of the object and thus provide a
possible scenario of globular cluster formation for collapsing
galaxies.
\\
Recently
Puy and Signore (1997), Bougleux and Galli (1997) showed that during
the early stages of gravitational collapse, for some collapsing masses, 
$HD$ molecules were the main cooling agent when the collapsing
protostructure had a temperature of about 200 Kelvins. In our last papers 
(Puy and Signore 1997, 1998) we have numerically calculated the cooling
functions for $H_2$, $HD$ and $LiH$ molecules by considering the
excitation of the twenty first rotational levels. $H_2$ is an
homonuclear molecule, so radiative transitions within the
lowest electronic state involve only quadrupole transitions, $\Delta
J=2$. Between two rotational levels, $\Delta J=J+2-J$ the energy of
the transition is given by
\begin{equation}
E_{J+2,J} \, = \, 2 k B_r (2J +3) \, = \, k T_{J+2,J}
\end{equation}
where $T_{J+2,J}$ is the temperature of the quadrupolar transition $J+2 \,
\rightarrow \, J$, $B_r$ the rotational constant (for $H_2$,
$B_r=85.33$ Kelvins) and $k$ the Boltzmann constant. Due
to the non-zero dipole moment the transitions involve dipolar
transitions, $\Delta J=1$. Thus the energy of the transition $J
\rightarrow J+1$ is given by:
\begin{equation}
E_{J+1,J} \, = \, 2 k B_r (J+1) \, = \, k T_{J+1,J}
\end{equation}
$T_{J+1,J}$ is the temperature of the dipolar transition $J+1 \, 
\rightarrow \, J$, $B_r = 64.3$ Kelvins for $HD$ and $B_r=10.7$
Kelvins for $LiH$. 
\\
In this paper we consider a collapsing molecular
protocloud made up only of $H_2$ and $HD$ molecules. We do not
consider collisions with $H$ atoms because we consider only the
central part of the cloud where the molecules are present. In this
picture $H$ is less heavier and is confined in the circumcloud. 
Moreover, we do not
consider the evolution of abundances. Puy \&
Signore (1998) showed that the evolution of $H_2$ and $HD$
abundance was low, and abundance of $LiH$ was negligible (see also 
Stancil et al. 1996). 
Thus, the constant ratio $\eta \, = \, n_{H_2}/
n_{HD}$ (where $n_{H_2}$ and $n_{HD}$ are respectively the density of 
$H_2$ and the density of $HD$) around the primordial ratio 
$$
\eta_{prim.} \, =\, n_{H_2} / n_{HD} \, \sim \, 1430
$$
(in the framework of the Standard Big Bang nucleosynthesis, see Puy et al 1993). 
\\
In section 2, we recall the primordial chemistry. In section 3, with these approximations, we analytically calculate the 
population of the ground state level and first state level in order to
evaluate the molecular cooling and molecular heating due to
$HD$ and $H_2$ molecules. In section 4, we estimate the
molecular cooling during the collapse of protoclouds. Then, in section
5, the possibility of thermal instability during the early phase of
gravitational collapse is discussed. Finally, in section 6, we 
suggest that this study could be extended to 
$CO$ molecules which have been recently
detected at redshifts $z>4$ (Ohta et al 1996, Omont et al 1996, 
Guilloteau et al 1997).
\section{Primordial chemistry}
In standard Big Bang theory, chemistry took place around the epoch of
recombination. At z$\sim$ 1000 the chemical species essentially were 
(see Signore \& Puy 1999):
$$
H, H^+, D, D^+, He, Li
$$
As the Universe expanded and cooled, different routes led to molecular
formation. Many authors have described in detail primordial molecular formation
and evolution; in particular, Puy et al. (1993) have simultaneously solved the
coupled chemical equations and the density and temperature evolution equations.
Finally with initial conditions given by the Standard Big Bang Nucleosynthesis 
(Signore \& Puy 1999), we obtain the following abundances (see Puy \& Signore
1999):
$$
H2/H \sim 8.4 \times 10^{-5}
$$
$$
HD/H \sim 10^{-5}
$$
$$
LiH/H \sim 10^{-17}
$$
In the following due to the very low abundance of $LiH$ we will consider only the
primordial molecules $H_2$ and $HD$.
\section{Analytic calculations of cooling and heating functions of
$H_2$ and $HD$ molecules}
We consider here only the transition between the ground state and the
first state. Due to the existence of a permanent dipolar moment for
$HD$ ($\mu_{HD} \, = \, 8.3 \, 10^{-4}$ Debye, Abgrall et al 1982),
the transition concerning $HD$ molecule is a dipolar transition
whereas for $H_2$ molecule the transition is quadrupolar (because it
has no permanent dipolar moment).
\subsection{HD molecule}
The excitation of the ground state is radiative or collisional, the
equations of evolution concerning the populations ($X_o^{HD}$ for the
ground state $J=0$, $X_1^{HD}$ for the first excited state $J=1$) 
are given by:

\begin{equation}
\frac{dX_1^{HD}}{dt} \,= \, 
\Bigl[ C_{0,1}^{HD} + B_{0,1}^{HD} u^{HD}_{0,1} 
\Bigr ] X_o^{HD} - 
\Bigl [ A_{1,0}^{HD} +C_{1,0}^{HD} + B_{1,0}^{HD} u^{HD}_{1,0}
\Bigr ] X_1^{HD} \nonumber
\end{equation}

\begin{equation}
\frac{dX_o^{HD}}{dt} \,= \,  
\Bigl [ A_{1,0}^{HD} +C_{1,0}^{HD} + B_{1,0}^{HD} u^{HD}_{1,0}
\Bigr ] X_1^{HD} -
\Bigl [ C_{0,1}^{HD} + B_{0,1}^{HD} u^{HD}_{0,1} 
\Bigr ] X_o^{HD} \nonumber 
\end{equation}
where $C_{0,1}^{HD}$ and $C_{1,0}^{HD}$ are the collisional
coefficients characterizing the collisions between $HD$ and $H_2$. 
In the approximation
of Maxwell-Boltzmann distribution we have:
$$
C_{0.1}^{HD} \, = \, 3 C_{1,0}^{HD} exp\Bigl ( -\frac{T_{1,0}^{HD}}{T_m}
\Bigr )
$$
with $T_{1,0}^{HD}=T_{0,1}^{HD}$ the transition temperature between
$J=0 \, \rightarrow \, J=1$, $T_m$ the temperature of the
matter inside the collapsing cloud, $B^{HD}_{0,1}$ and
$B^{HD}_{1,0}$ are the Einstein coefficients. The radiative density is
$$
u^{HD}_{1,0} = u^{HD}_{0,1} \, = \,
\frac{8 \pi h (\nu_{10}^{HD})^3}{c^3}
.\frac{1}{exp \Bigl (\frac{T_{1,0}^{HD}}{T_r}
\Bigr ) - 1}
$$
where $\nu_{1,0}^{HD}$ and $T_{1,0}^{HD}$ are the corresponding 
frequency and temperature, $h$ the Planck constant, $c$ the
velocity of the light and $T_r$ the temperature of the background radiation.
$$ 
A_{1,0}^{HD} \, = \, \frac{512 \, \pi^4 \, k^3}{3 h^4
c^3}.\frac{B_{HD}^3 
\mu_{HD}^2}{3}
$$
characterizes the Einstein coefficient for the spontaneous emission (see Kutner
1984) and $B_{HD} \, = \, 64.3$ Kelvins the rotational constant (see Herzberg
1950). 
\\
We have the following normalisation for the populations: 
$$
X_o^{HD} + X_1^{HD} \, = \, 1
$$
Moreover we consider a
quasi-instantaneous transition so 
$$
\frac{d X_o^{HD}}{dt} \, = \, \frac{d X_1^{HD}}{dt} \, = \, 0
$$
Thus from these approximations we find for the populations $X_1^{HD}$
and $X_o^{HD}$:
\begin{equation}
X_1^{HD} \, = \, 
\frac{B_{0,1}^{HD} u_{0,1}^{HD} + C_{0,1}^{HD}}
{A_{1,0}^{HD} + u_{1,0}^{HD} ( B_{0.1}^{HD} + B_{1,0}^{HD} )
+ C_{0,1}^{HD} + C_{1,0}^{HD} }
\end{equation}
and
\begin{equation}
X_o^{HD} \, = \, 
\frac{A_{1,0}^{HD} + B_{1,0}^{HD} u_{1,0}^{HD} + C_{1,0}^{HD}}
{A_{1,0}^{HD} + u_{1,0}^{HD} ( B_{0.1}^{HD} + B_{1,0}^{HD} )
+ C_{0,1}^{HD} + C_{1,0}^{HD} }
\end{equation}
The probability of collisional de-excitation is given by
\begin{equation}
P_c^{HD} \, = \, \frac{C_{1,0}^{HD}}{C_{1,0}^{HD} + A_{1,0}^{HD} + 
B_{1,0}^{HD} u_{1,0}^{HD}}
\end{equation}
and the probability of radiative de-excitation:
\begin{equation}
P_r^{HD} \, = \, \frac{A_{1,0}^{HD}+B_{1,0}^{HD} u_{1,0}^{HD}}
{C_{1,0}^{HD} + A_{1,0}^{HD} + B_{1,0}^{HD} u_{1,0}^{HD}}
\end{equation}
Thus, we can calculate the molecular cooling (collisional excitation
followed by a radiative de-excitation):
\begin{equation}
\Lambda_{HD} \, = \, n_{HD} X_o^{HD} C_{0.1}^{HD} P_r^{HD}
E_{1,0}^{HD}
\end{equation}
where $E_{1,0}^{HD} \, = \, k T_{1,0}^{HD}$ is the energy for the
transition $J=0 \, \rightarrow \, J=1$.
\\
The molecular heating (radiative excitation followed by collisional
de-excitation) is :
\begin{equation}
\Gamma^{HD} \, = \, n_{HD} X_o B_{0,1}^{HD} u_{0,1}^{HD} P_c^{HD}
E_{1,0}^{HD} 
\end{equation}

If we take into account the following relation between the Einstein
coefficients:
$$
B_{0,1}^{HD} \, = \, 3 B_{1,0}^{HD} \ \ {\rm and} \ \ 
B_{1,0}^{HD} \, = \, \frac{c^3}{8 \pi h (\nu_{1,0}^{HD})^3} A_{1,0}^{HD}
$$
then we deduce for the molecular cooling
\begin{equation}
\Lambda_{HD} \, = \,
\frac{3 C_{1,0}^{HD} exp(-\frac{T_{1,0}^{HD}}{T_m})
exp(\frac{T_{1,0}^{HD}}{T_r}) h \nu_{1,0}^{HD} n_{HD} A_{1,0}^{HD}}
{A_{1,0}^{HD}\Bigl [ 3+exp(\frac{T_{1,0}^{HD}}{T_r})\Bigr ] +
C_{1,0}^{HD} \Bigl [ 1 + 3 exp(-\frac{T_{1,0}^{HD}}{T_m}) \Bigr ]
\Bigl [exp(\frac{T_{1,0}^{HD}}{T_r}) -1 \Bigr ]}
\end{equation}
and for the molecular heating
\begin{equation}
\Gamma_{HD} \, = \,
\frac{3 C_{1,0}^{HD} h \nu_{1,0}^{HD} n_{HD} A_{1,0}^{HD}}
{A_{1,0}^{HD}\Bigl [ 3+exp(\frac{T_{1,0}^{HD}}{T_r})\Bigr ] +
C_{1,0}^{HD} \Bigl [ 1 + 3 exp(-\frac{T_{1,0}^{HD}}{T_m}) \Bigr ]
\Bigl [exp(\frac{T_{1,0}^{HD}}{T_r}) -1 \Bigr ]}
\end{equation}
Notice that calculating the ratio molecular cooling to molecular
heating we obtain the following simple expression
\begin{equation}
\eta ^{HD} \, = \, \frac{\Lambda ^{HD}}{\Gamma ^{HD}}
\, = \, 
exp \Bigl [ \frac{T_{1,0}^{HD}(T_m - T_r)}{T_r T_m} \Bigr ]
\end{equation}
This relation gives us an important information concerning the thermal
influence of the thermal function $\Psi ^{HD} \, = \, \Gamma ^{HD} - 
\Lambda ^{HD}$ during a gravitational collapse. When the matter 
temperature $T_m$ is higher than the radiative temperature, we
have $\eta ^{HD} > 1$, i.e. $\Lambda ^{HD} > \Gamma ^{HD}$ so molecules
cool the collapsing cloud. Thus during a gravitational collapse, the 
matter temperature of the collapsing cloud is higher than the radiative
temperature of the cosmological background (let us recall that 
in our case we do not consider some external sources like quasars, 
first stars etc...).
\\
Thus in this study, we consider only the molecular cooling which is
the dominant term in the cooling function as we have shown for
a collapsing cloud (Puy \& Signore 1996). Moreover if we compare the
temperature of the dipolar transition for $HD$, $J=0 \, \rightarrow \,
J=1$, $T_{10}^{HD}= 128.6$ K with the temperature of the cosmological
radiation between the redshifts $[5,20]$, i.e. $T_r \, \in \,
[13.5,54]$, we can approximate:
$$
3 +exp \Bigl (\frac{T_{10}^{HD}}{T_r} \Bigr ) 
\, \sim \, exp \Bigl ( \frac{T_{10}^{HD}}{T_r} \Bigr ) 
$$ 
and
$$
exp \Bigl ( \frac{T_{10}^{HD}}{T_r} \Bigr ) 
- 1 \, \sim \, exp \Bigl ( \frac{T_{10}^{HD}}{T_r} \Bigr ) 
$$
Therefore, with these approximations, we obtain for the cooling
function of $HD$ (Eq. 9) the expression:
\begin{equation}
\Lambda_{HD} \, \sim \,
\frac{3 C_{1,0}^{HD} exp(-\frac{T_{1,0}^{HD}}{T_m})
 h \nu_{1,0}^{HD} n_{HD} A_{1,0}^{HD}}
{A_{1,0}^{HD} +
C_{1,0}^{HD} \Bigl [ 1 + 3 exp(-\frac{T_{1,0}^{HD}}{T_m}) \Bigr ]}
\end{equation} 

\subsection{$H_2$ molecule}

Concerning the $H_2$ molecules, the analysis is the same than that for the $HD$
molecules with the difference that the studied transitions are
quadrupolar. Because of the increased number of charged particle regions
with molecules ($H_2$, $HD$...), the heating mechanism is different
from and more complicated than that of atomic regions. We focus 
our calculations only on the first transition 
$J=0 \, \rightarrow \, J=2$. The other 
transitions are negligible because we consider the early phase of
gravitational collapse where the matter temperature is below 200 Kelvins which 
is much lower than the transition temperatures for $H_2$ molecules
(see Table 1). The 
Einstein coefficients for the spontaneous emission and for the first
transition are given by (see Kutner 1984):
$$
A_{2,0}^{H_2} \, = \, 2.44 \times 10^{-11} \ {\rm s}
$$
The relations between the Einstein coefficients are
$$
B_{0,2}^{H_2} \, = \, 5 B_{2,0}^{H_2} \ \ {\rm and} \ \ 
B_{2,0}^{H_2} \, = \, \frac{c^3}{8 \pi (\nu_{2,0}^{H_2})^3}A_{2,0}^{H_2}
$$
and the radiative density for the transition is:
$$
u_{2,0}^{H_2} \, = \, u_{0,2}^{H_2} \, = \, \frac{8 \pi h
 (\nu_{2,0}^{HD})^3}{c^3} . \frac{1}{exp \Bigl (
\frac{T_{0,2}^{H_2}}{T_r}-1 \Bigr ) }
$$
and for the collisional coefficients  in the Maxwell-Bolztmann
distribution approximation:
$$
5 C_{2,0}^{H_2} \, = \, C_{0,2}^{H_2} exp \Bigl (
\frac{T_{2,0}^{H_2}}{T_m}
\Bigr )
$$
by analogy with the $HD$ molecules and with the same levels of approximations
$$
exp\Bigl ( \frac{T_{2,0}^{H_2}}{T_r} \Bigr ) - 1  \, \sim \, 
exp\Bigl ( \frac{T_{2,0}^{H_2}}{T_r} \Bigr )
$$
and
$$ 
5 + exp\Bigl ( \frac{T_{2,0}^{H_2}}{T_r} \Bigr ) \, \sim \, 
exp\Bigl ( \frac{T_{2,0}^{H_2}}{T_r} \Bigr )
$$
Thus, from the Eqs (5) and (6), we find for the populations
$X_o^{H_2}$ and $X_1^{H_2}$:
$$
X_o^{H_2} \, \sim \, 
\frac{A_{2,0}^{H_2} +C_{2,0}^{H_2}}{A_{2,0}^{H_2} +C_{2,0}^{H_2}
\Bigl (1 + exp ( \frac{T_{2,0}^{H_2}}{T_m} ) \Bigr )
}
$$
$$
X_{1}^{H_2} \, \sim \, 
\frac{
5C_{2,0}^{H_2} exp ( \frac{T_{2,0}^{H_2}}{T_m} )}
{A_{2,0}^{H_2} +C_{2,0}^{H_2}
\Bigl (1 + exp ( \frac{T_{2,0}^{H_2}}{T_m} ) \Bigr )
}
$$
The probability of radiative de-excitation 
becomes
\begin{equation}
P_r^{H_2} \, \sim \, \frac{A_{2,0}^{H_2}}{A_{2,0}^{H_2}+C_{2,0}^{H_2}}
\end{equation}
while the molecular cooling for the $H_2$ molecule is given by:
\begin{equation}
\Lambda_{H_2} \, \sim \, 
\frac{
5 n_{H_2} C_{2,0}^{H_2} A_{2,0}^{H_2} h \nu _{2,0}^{H_2} 
exp(-\frac{T_{2,0}^{H_2}}{T_m})
}
{A_{2,0}^{H_2}+C_{2,0}^{H_2} \Bigl [
1+5 exp(-\frac{T_{2,0}^{H_2}}{T_m}) \Bigr ]
}
\end{equation}

\subsection{Collision rates}
The collisional coefficients are given by the expressions:

\begin{equation}
C_{1,0}^{HD} \, = \, < \sigma . v^{HD} > \, n_{H_2} 
\end{equation}
\begin{equation}
C_{2,0}^{H_2} \, = \, < \sigma . v^{H_2} > \, n_{H_2} \\
\end{equation}
where  $\sigma \, \sim \, 1 \AA ^2$ is the collisional cross section and
$n_{H_2}$ the density of $H_2$; $v_{HD}$ and $v_{H_2}$ are
respectively the velocity of $HD$ and $H_2$ given by
$$
v_{HD} \, = \, v_{H_2} \, = \, \sqrt{\frac{3 k T_m}{2 m_H}}
$$
where $m_H$ is the hydrogen mass. Notice that, in our
model, $H_2$ molecules are more abundant than $HD$ molecules, so $H_2$
molecules are the main collisional species.
\subsection{Total cooling function}
The total cooling function is given by 
\begin{equation}
\Lambda_{Total} \, = \, \Lambda_{HD} + \Lambda_{H_2} \, = \, 
\Lambda_{HD} (1 + \xi_{H_2})
\end{equation}
where 
$$
\xi_{H_2} \, = \, \frac{\Lambda_{H_2}}{\Lambda_{HD}}$$ 
is the ratio
between the $H_2$ and $HD$ cooling functions. With the help of the numerical
values of the Table 1, we deduce:
\begin{equation}
\Lambda_{HD} \, \sim \, 2.66 \times 10^{-21} n^2_{HD} \, 
exp(-\frac{128.6}{T_m})
\frac{\sqrt{T_m}}{1 + n_{HD} \sqrt{T_m} \Bigl [ 1 + 3 
exp(-\frac{128.6}{T_m}) \Bigr] } 
\end{equation} 
\begin{equation}
\Lambda_{H_2} \, \sim \, 1.23 \times 10^{-20} n^2_{HD} \, exp(-\frac{512}{T_m})
\frac{\sqrt{T_m}}{5 \times 10^{-4} + n_{HD} \sqrt{T_m} \Bigl [ 1 + 5 
exp(-\frac{512}{T_m}) \Bigr] } 
\end{equation} 
and for the ratio:
\begin{equation}
\xi_{H_2} \, \sim  \, 
4.63 \, e^{-\frac{383.4}{T_m}} \, \frac{
1+\sqrt{T_m}n_{HD} [ 1 + 3 exp(-\frac{128.6}{T_m})]}
{5 \times 10^{-4} +\sqrt{T_m} n_{HD} [ 1 + 5
exp(-\frac{512}{T_m})]}
\end{equation}
\section{Molecular functions of a collapsing protocloud}
We study a homologous model of spherical collapse of mass
$M$ similar to the model adopted in Lahav (1986) and Puy \& Signore
(1996) in which we only consider $H_2$ and $HD$ molecules. 
\\
We have seen (Puy \& Signore 1997) from a numerical integration of 
the cooling functions where twenty rotational levels were considered 
that molecular cooling is important. Moreover we have shown that below 
200 Kelvins the main cooling agent is the $HD$ molecule. The aim 
of this paper is to analyse, through our approximations (where the
only two first excited levels are considered), the evolution of the cooling and
 the potentiality of thermal instability. 
\\
First, let us recall the equations governing dynamics of a collapsing
protocloud: 
\begin{equation} 
\frac{dT_m}{dt} \, = \, 
-2 \frac{T}{r}.\frac{dr}{dt} - \frac{2}{3n k}\Lambda_{Total}
\end{equation}
for the evolution of the matter temperature $T_m$,
\begin{equation}
\frac{d^2 r}{dt^2} \, = \, \frac{5k T_m}{2 m_H r} - 
\frac{G M}{r^2}
\end{equation}
for the evolution of the radius $r$ of the collapsing cloud and
\begin{equation}
\frac{dn}{dt} \, = \, -3 \frac{n}{r}.\frac{dr}{dt}
\end{equation}
for the evolution of the matter density $n$.
\\
At the beginning of the gravitational
collapse the matter temperature increases, then due to the very
important efficiency of the molecular cooling, the temperature
decreases. We consider here this {\it transition regime} i.e. the point where
the temperature curve has a horizontal asymptot (see Puy \& Signore 
1997): 
$$
\frac{dT_m}{dt} \, = \, 0
$$
and where the cooling due to molecules can exceed the heating
resulting from the collapse. 
From the equation (22) of the evolution of the matter
temperature, we deduce
$$
-2\frac{T_m}{r}.\frac{dr}{dt} \, = \, \frac{2}{3nk}\, \Lambda_{Total}
$$
Note that the total molecular cooling is negative
($\Lambda_{Total} < 0$). Moreover, because we consider the early phase of the
gravitational collapse, we can approximate to a free fall
collapse and so the pressure term, $5 k T_m/2 m_H r$, is
negligible in the equation (23). 
Thus, the evolution equation of the radius can be written:
$$
\frac{d^2r}{dt^2} \, = - \, \frac{G M}{r^2}
$$
which leads to 
\begin{equation}
\frac{dr}{dt} \, = \, \frac{GM}{r}
\end{equation}
Moreover, because we know that the evolution of the density is: 
\begin{equation}
n \, = \, n_o (\frac{r}{r_o})^3
\end{equation}
where $n_o$ and $r_o$ are respectively the density and the radius at
the beginning of the collapse. Thus we conclude that 
\begin{equation}
\Lambda_{Total} \, = \, \frac{3 n_o k T_m r_o^3}{r^5} GM
\end{equation}						
We have seen that the regime of temperature that we consider is below
200 Kelvins. Moreover, the abundances of $HD$ and $H_2$ remain roughly
constant during the early phase of the gravitational collapse 
(Puy \& Signore 1997, 1998). In this case, the value of the
density of $HD$ at the beginning of the collapse is
$n_{HD} \, \sim \, 6 \times 10^{-8}$ cm$^{-3}$. The temperature of the
matter is below $200$ Kelvins. Thus we can approximate the following
terms as follows:
$$
1 + n_{HD} \sqrt{T_m} \Bigl [ 1+3exp(-\frac{128.6}{T_m})\Bigr] \, \sim \, 1
$$ 

$$ 
5 \times 10^{-4} + n_{HD} \sqrt{T_m} 
 \Bigl [ 1+5exp(-\frac{512}{T_m})\Bigr ]
\, \sim \, 5 \times 10^{-4}
$$
These approximations lead to the following simplified expressions for
the molecular cooling 
\begin{equation}
\Lambda_{HD} \, \sim \, \Lambda_o \, exp(-\frac{T_o}{T_m})
\, n^2_{HD} \, \sqrt{T_m}
\end{equation}
where $\Lambda_o$; $T_o$ and $\Delta T_o$ are constants:
$$
\Lambda_o \, = \, 2.66 \times 10^{-21} \ {\rm erg \ cm^{-3} \ K^{-1/2} \
s^{-1}}
$$
$$
T_o \, = \, T_{1,0}^{HD} \, = \, 128.6 \ {\rm K}
$$
$$
\Delta T_o \, = T_{2,0}^{H_2} - T_{1,0}^{HD} \, = \, 383.4 \ {\rm K}
$$
From our homologous model of gravitational collapse, the
density of matter is the density of the species which is the most
abundant so $n=n_{H_2}$ which gives 
\begin{equation}
n_{HD} \, = \, \frac{n_{H_2}}{\eta_{Prim.}} \, = \,
\frac{n}{\eta_{Prim.}} \, = \, \frac{n_o }{\eta_{Prim.}}.
\Bigl ( \frac{r_o}{r} \Bigr )^3 
\end{equation}
Finally we find for the total cooling
\begin{equation}
\Lambda_{Total} \, \sim \, \Lambda_o exp(-\frac{T_o}{T_m}) \frac{n_o^2
r_o^6}{\eta_{Prim.}^2 r^6} \sqrt{T_m} \Bigl [
1 + \xi_o exp(-\frac{\Delta
T_o}{T_m})
\Bigr ]
\end{equation}
With $\xi_o = 9260$, 
if we take for the mass of the cloud the following approximation: 
$$
M \, \sim \, 2 m_H n_o r^3_o
$$
from equations (28) and (32) we deduce for the the radius
\begin{equation}
r \, \sim \, \kappa \frac{exp(-T_o/T_m) \Bigl [ 1 + \xi_o
exp(-\Delta T_o / T_m) \Bigr ] }{\sqrt{T_m}} 
\end{equation}
where the constant 
$$
\kappa= \frac{\Lambda_o}{6 \eta_{Prim.}^2 k G m_H} \, \sim \, 
1.4 \times 10^{22}
$$
Finally we obtain for the total molecular cooling
\begin{equation}
\Lambda _{Total} \, = \, \delta \, T_m^{7/2} \, 
\frac{exp(5T_o/T_m)}{\Bigl[ 1 + \xi_o exp(-\Delta T_o / T_m)
\Bigr ]^5}
\end{equation}
with 
$$
\delta \, = \, \alpha \, M^2 \ {\rm with} \ \alpha \, = \, 
\frac{3 k G}{ 2 m_H \kappa^5}
$$
\section{Thermal Instability}
We have learned much about thermal instability in the last 30 years since
the appearance of the work of Field (1965). Many studies have
focused on the problem of thermal instabilitie in different situations:
formation of galaxies (Sofue 1969), expanding medium (Kondo 1970),
formation of protostars (Stein et al. 1972), formation of globular
clusters (Fall \& Rees 1985), cooling flows (David et al.
1988) and quasar clouds (Mathews \& Doane 1990). 
\\
The correct instability criterion can be derived physically as
follows. In an infinite, uniform, static medium of density $n_o$ and 
$T_o$, the continuity and force equations are automatically
satisfied. If we introduce a perturbation of density and temperature
such that some thermodynamics variable (pressure, temperature...) is
held constant, the entropy of the material $\cal{S}$ will change by an amount
$\delta {\cal{S}}$, and the heat-loss function, by an amount $\delta 
\Psi=\Gamma - \Lambda$. From the equation 
$$
\delta \Psi \, = \, -T d(\delta {\cal{S}} )
$$
there is instability only if:
$$
\Bigl ( \frac{\delta \Psi}{\delta {\cal{S}}} \Bigr ) \, > \, 0
$$
We know that $T d{\cal{S}} = C_p dT$ in an 
isobaric perturbation. The corresponding inequality can be written: 
$$
\Big ( \frac{\delta \Psi}{\delta T} \Big )_p 
\, = \, 
\Big ( \frac{\delta \Psi}{\delta T} \Big )_n
+
\frac{\partial n}{\partial T} \,
\Big ( \frac{\delta \Psi}{\delta n} \Big )_T
\, < \, 0
$$
which characterizes the Field's criterion. The criterion involves
constant $P$ because small blobs tend to maintain pressure equilibrium
with their surroundings when heating and cooling get out of balance. 
In our case (i.e. at the transition regime), a thermal
instability could spontaneously be developped if the
Field's Criterion is verified:
\begin{equation}
\frac{\partial \, ln \Lambda }{\partial \, ln T_m} \, < \, 0
\end{equation} 
In our case the logaritmic differential of the total molecular cooling
function is given by 
\begin{equation}
\frac{\partial \, ln \Lambda}{\partial \, ln T_m} \, = \, 
\frac{
7T_m +7T_m \xi_o e^{-\frac{\Delta T_o}{T_m}}-10 T_o -
10T_o \xi_o e^{-\frac{\Delta T_o}{  T_m}} - 10 \xi_o \Delta T_o 
e^{-\frac{\Delta T_o }{ T_m}}
}
{2 + 2 \xi_o e^{-\frac{\Delta T_o}{ T_m}}}
\end{equation}
Therefore, the criterion is independant of 
the mass $M$ of the cloud at the transition regime. 
In figure 1, we have plotted the curve $y(T_m) \, = \, \partial \, ln
\Lambda / \partial \, ln T_m $. It shows that the {\it
Field's criterion} is always verified. 
\\
We conclude that a thermal instability is possible at the {\it 
transition regime}, i.e. at $dT_m/dt =0$, when cooling (due to molecules) 
exceeds heating (due essentially to the collapse). Thus by maintening
the same pressure in its surroundings, such a blob would get cooler
and cooler and denser and denser until it could separate into
miniblobs. This possibility is very interesting and could give a
scenario of formation of primordial clouds. However, a quantitative
study is necessary to evaluate the order of magnitude of the
mass and the size of the clouds. This last point can be crucial
particularly concerning the composition of the molecular clouds at high
redshift such as the Lyman-$\alpha$ absorbers in order to
evaluate the possible signatures of the primordial clouds.


\begin{figure}[t!] 
\centerline{\epsfig{file=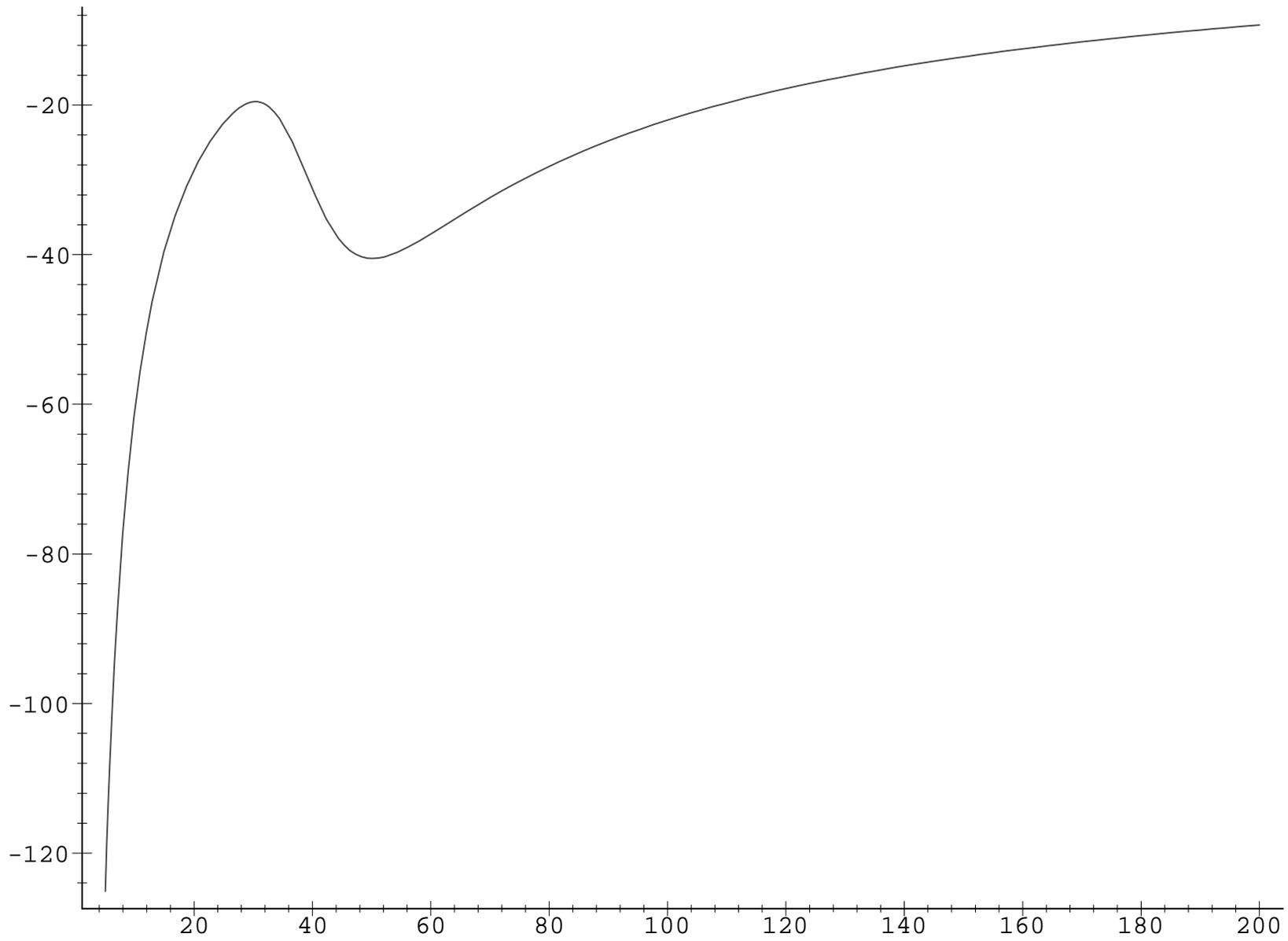}}
\caption{\small{$\partial ln \Lambda_{Total}/
\partial ln T_m$ in y-axis and temperature of the matter in x-axis
(in Kelvins) }}
\vspace*{8pt}
\label{fig1}
\end{figure}

\section{On $CO$ molecule}
In the very early Universe the components of the primordial molecular
medium are formed from the elements of the primordial
nucleosynthesis. Later, as soon as stellar processes occur in 
proto-objects, other molecular species appeared such as $CO$, $CI$,
$HCN$. Therefore, their line emission can be seen as an excellent probe
of physical conditions of very high redshift objects. Among them, the
diatomic molecule $CO$ has the strongest lines at radio wavelengths, 
the rotational transitions being spaced by 115 Ghz in the rest frame
at $z=0$ (this separation, at a redshift $z$, becomes compressed by a
factor $1/(1+z)$). Moreover, the carbon monoxide
line emission had already been detected in four high redshift objects:
\begin{enumerate}
\item $CO(3-2)$, $CO(4-3)$ and $CO(6-5)$ in the ultraluminous IR galaxy
F10215+4724 at the redshift $z=2.28$ (Brown and Van den Bout 1992,
Solomon et al 1992).
\item $CO(3-2)$, $CO(4-3)$, $CO(5-4)$ and $CO(7-6)$ in the cloverleaf
quasar at the redshift $z=2.56$ (Barvainis et al 1994, Barvainis et al
1997).
\item $CO(5-4)$ in the quasar BR1202-0725 at the redshift $z=4.69$
(Ohta et al 1996, Omont et al 1996).
\item $CO(5-4)$ in the quasar BRI1335-0418 at the redshift $z=4.41$ 
(Guilloteau et al 1997). 
\end{enumerate}
One must also add that, at least for the first three objects, their
detectability in molecular transitions may be due to high emissivity
of the gas and above all magnification from gravitational lensing
rather than an extremely large mass of gas. Moreover, there is no a
priori evidence of gravitational lensing for BRI1335-0418. 
Therefore, if this absence of strong gravitational magnification is
confirmed, the fourth object could have a large mass of molecular gas
which could be associated with the presence of some huge
starburst. 
However, if there is no doubt about the reality of the $CO$ detection
in these objects, it is more difficult to definitively determine the
$CO$ to $H_2$ conversion factor in these objects.
\\
Let us also remark that, here, we study {\it protostructures} in the
mass range $10^8-10^{10}$ M$_\odot$ i.e. in the mass range of dwarf
galaxies. Recently, Verter \& Hodge (1995) estimated the $CO$ to
$H_2$ conversion factor for the very low metallicity dwarf irregular
galaxy GR8. They found $X > 80 \times 10^{20}$ molecules cm$^{-2}$ 
(Km s$^{-1}$)$^{-1}$ 
which is a factor 30 larger than for our Galaxy ($X=N(H_2)/W_{CO}$
where $N(H_2)$ is the column density of molecular hydrogen gas and
$W_{CO}$ is the integration of $CO$ antenna temperature over the
velocity profile).
\\
Therefore, from these estimations, the calculations of molecular
functions could, in principle, be extended to the $CO$ molecules for
collapsing protoclouds at $z<5$. The $CO$ molecules could even be a
better cooling agent than $H_2$ and $HD$ for some temperature
ranges. But, in any case -and in particular for the discussion of
thermal instability at the {\it transition regime}- the corresponding
total cooling function must be numerically calculated because the
excitation of many rotational levels must be taken into account. 
This extended study is beyond the scope of our paper. 
\section*{acknowledgements}
The authors gratefully acknowledge St\'ephane Charlot, Philippe
Jetzer, Lukas Grenacher, Bianca Melchiorri-Olivo and Francesco Melchiorri 
for valuable discussions on this field. 
Part of the work of D. Puy has been conducted under the auspices 
of the {\it D$^r$ Tomalla Foundation} and Swiss National Science Foundation.

\clearpage
\section*{References}
{\footnotesize
\noindent
Abgrall H., Roueff E., Viala Y. 1982 A\&AS 50, 505
\\
Barvainis R., Tacconi L., Alloin D., Antonucci R.,
Coleman P. 1994 Nature 371, 586
\\
Barvainis R., Maloney P., Alloin D.,
Antonucci R. 1997 ApJ 484, 695
\\
Bodenheimer P. 1968 ApJ 153, 483
\\
Bougleux E., Galli D. 1997 MNRAS 288, 638
\\
Brown R. L., Van den Bout P.A. 1992 ApJ 397, L19
\\
Dalgarno A., R.A. Mc Cray 1972, Ann. Rev. Astr. and Ap. 10, 375
\\
David L.P., Bregman J.N., c. Gregory Seab 1988 ApJ 329, 66
\\
Fall S.M., Rees M.J. 1985 ApJ 298, 18
\\
Field G. B. 1965 ApJ 142, 531
\\
Guilloteau S., Omont A., Mc Mahon R.G.,
Cox P., Petitjean P. 1997 A\&A 328, L1
\\
Herzberg G. 1950 in {\it Spectra of
diatomic molecules} Van Nostrand eds
\\
Hutchins J.B. 1976 ApJ 205, 103
\\
Kondo M. PASJ 22, 13   
\\
Kutner M.L. 1984 Fund. Cosm. Phys. 9, 233
\\
Lahav O. 1986 MNRAS 220, 259
\\
Mathews W. G., Doane J. S. 1990 ApJ 352, 423
\\
Ohta K., Yamada T., Nakanishi K., Kokno K.,
Akiyama M., Kawabe R. 1996 Nature 382, 426
\\
Omont A., Petitjean P., Guilloteau S., Mc
Mahon R.G., Solomon P.M. Pecontal E. 1996 Nature 382, 428
\\
Palla F., Salpeter E.E., Stahler S.W. 1983
ApJ 271, 632
\\
Puy D., Alecian G., Lebourlot J., L\'eorat J.,
Pineau des Forets G. 1993 A\&A 267, 337 
\\
Puy D., Signore M., 1996, A\&A 305, 371
\\
Puy D., Signore M., 1997, New Astron. 2, 299
\\
Puy D., Signore M., 1998, New Astron. 3, 27
\\
Puy D., Signore M., 1999, New Astron. Rev. {\it in press}
\\
Shu F.H. 1997 in {\it Molecules in
Astrophysics: Probes and Processes} pp19, IAU 1997 Van Dishoeck ed.
\\
Sofue Y. 1969 PASJ 21, 211
\\
Solomon P.M., Downes D., Radfort
S.J.E. 1992 ApJ 398, L29
\\
Stancil P.C., Lepp S., Dalgarno A. 1996 ApJ
458, 401
\\
Stein R.F., Mc Cray R., Schwarz J. 1972 ApJ 177, L125
\\
Signore M., Puy D., 1999 New Astron. Rev. {\it in press}
\\
Tegmark M., Silk J., Rees M.J., Blanchard
A., Abel T., Palla F. 1997 ApJ 474,1.
\\
Verter F., Hodge P. 1995 ApJ 446, 616
\\
Villere K. R., Bodenheimer P. in {\it Astrochemistry} pp121, IAU 120, 
eds Vardia, S.P. Tarafdar, Reidel 1987
} 
\end{document}